\documentclass[arXiv,
]{actapoly}
\usepackage[utf8]{inputenc}

\usepackage{lineno}

\begin{document}

\title[Example of an Article with a Long Title]
{Interstellar gas motions around massive star formation regions in the nearby dwarf galaxy DDO\,43}

\correspondingauthor[Enikő Pichler]{Enikő Pichler}{first}{pichler.eniko@gmail.com}
\author[Bendegúz Koncz]{Bendegúz Koncz}{first,second}
\author[Krisztina É. Gabányi]{Krisztina É. Gabányi}{first,third,fourth}
\author[András Péter Joó]{András Péter Joó}{first}
\author[L. Viktor Tóth]{L. Viktor Tóth}
{first,second}
\institution{first}{ELTE Eötvös Loránd University, Department of Astronomy, Pázmány Péter sétány 1/A, H-1117 Budapest, Hungary}
\institution{second}{Doctoral School of Physics at University of Debrecen
4026 Debrecen, Bem tér 18/B
Hungary}
\institution{third}{HUN-REN--ELTE Extragalactic Astrophysics Research Group, ELTE Eötvös Loránd University, Pázmány Péter sétány 1/A, H-1117 Budapest, Hungary}
\institution{fourth}{Konkoly Observatory, HUN-REN Research Centre for Astronomy and Earth Sciences, Konkoly Thege Miklós út 15-17, H-1121 Budapest, Hungary}

\begin{abstract} Areas of massive star formation are strongly influenced by stellar winds and supernovae, therefore, enhanced turbulent flows are expected. We analyse high-quality Karl G. Jansky Very Large Array observations of the neutral hydrogen gas content of DDO\,43, a relatively nearby irregular dwarf galaxy. The line wings of neutral hydrogen spectral lines, which provide insights into local enhanced velocity dispersion, are investigated together with far-ultraviolet data, tracing emissions from massive star-forming regions. We find very weak correlations with both higher and lower velocity areas.

\end{abstract}

\keywords{line: profiles, turbulence, methods: data analysis, surveys, galaxies: ISM, radio lines: ISM, ultraviolet: stars}

\maketitle

\section{Introduction}

Stellar winds and supernovae strongly influence areas of massive star formation: in these regions, enhanced turbulent flows are expected. As neutral hydrogen (hereafter HI) is the most abundant element in the Universe, it is an effective tracer of the energy that star formation injects into the local gas. Turbulence has also been studied based on the observations of emission lines tracing warm ionised gas (e.g. \cite{Law_2022}), and the tracer of molecular gas, CO has also been extensively studied (e.g. \cite{Bacchini_2020}). For a detailed review of interstellar turbulence, see \cite{Elmegreen_2004}.

The turbulence-enhancing effect of stellar evolution and its role in regulating further star formation has been studied in several researches (e.g. \cite{Ianjamasimanana_2015}) based on observations of the HI. Our study is based specifically on the data and results of \cite{Elmegreen_2022} and \cite{Hunter_2021}, who have found no trace of the expected excess turbulence in HI gas around massive star-forming regions. They determined the kinetic energy density and velocity dispersion based on the HI emission maps, and then compared these with star formation rate densities calculated from far-ultraviolet observations. Their research concluded that there was little or no correlation between these values, which means that the feedback energy from the star formation is not observable in the local atomic gas. 

\begin{figure}[H]
	\centering
		\includegraphics[width=0.75\linewidth]{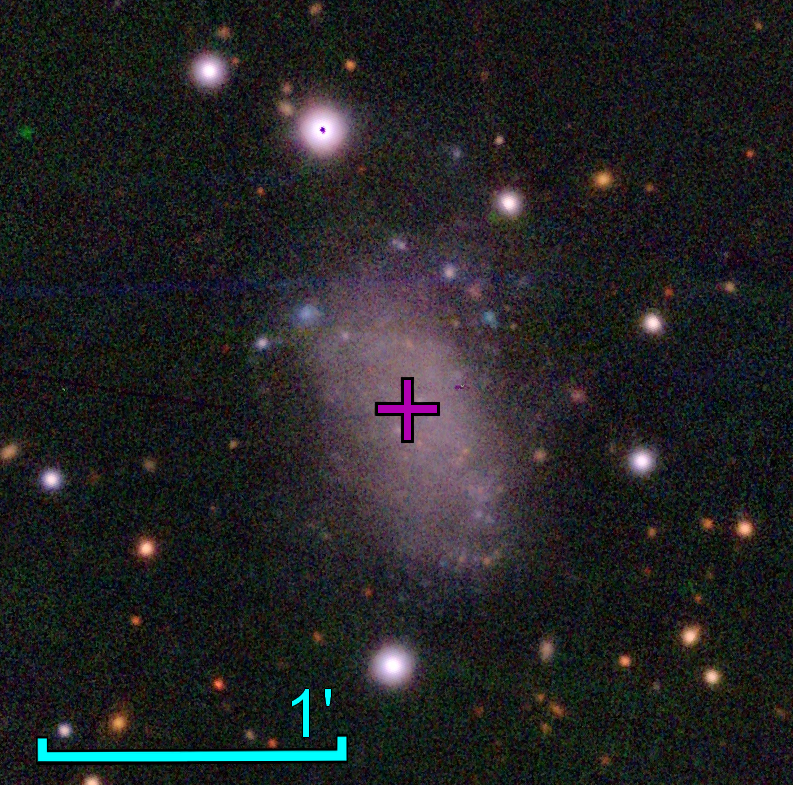}
    \hspace{5pt}
	\caption{The optical image of DDO\,43 in bands i, r, and g from Pan-STARRS \cite{chambers2019panstarrs1surveys,Flewelling_2020}. North is up, east is to the left. The purple cross refers to the centre of the galaxy, $RA= 07^\textrm{h}28^\textrm{m}17.724^\textrm{s}$, and $Dec=+40$\,$^{\circ}$ $46'11.36''$.}
	\label{fig:galaxy-optical}
\end{figure}

We describe a different method for studying this relationship between interstellar gas and star formation, looking into the line wings of HI spectra, that provide information on local phenomena. Our study examines the HI spectra in DDO\,43 (also known as UGC\,3860 or PGC\,21073, see Figure \ref{fig:galaxy-optical}), an irregular dwarf galaxy at a distance of 7.8 Mpc \cite{Karachentsev_2004}. The HI content was measured within the framework of the Local Irregulars That Trace Luminosity Extremes, The HI Nearby Galaxy Survey (LITTLE THINGS) \cite{Hunter_2012}, which provides high-quality 21 cm VLA observations. As one of the bases of our study, \cite{Hunter_2021} investigates galaxies from the LITTLE THINGS survey, one of its targets, DDO\,43 was chosen as the first galaxy to examine with our method. The compact size of DDO\,43 makes it relatively easy to study the full extent of the HI in the galaxy, but it is not too small to resolve. Its inclination of 42$\degree$ \cite{Simpson_2005} allows the study of structures in its disk. In addition, as a dwarf with a lower gravitational potential, other effects that regulate star formation should be easier to observe \cite{Hunter_2021}. With these criteria, DDO\,43 was chosen arbitrarily from the LITTLE THINGS sample.

The star formation rate (SFR) of DDO\,43 was calculated to be $3.7 \times 10^{-3}$ M$_\odot\textrm{yr}^{-1}$ from H$\alpha$ luminosity (which traces the ionised gas present around massive stars) \cite{Simpson_2005, Hunter_2004}. This rate counts as average among irregular dwarfs \cite{Simpson_2005}. In the far-ultraviolet (which is another tracer of star formation as the spectral energy distribution of young high-mass stars have their peak in the FUV), DDO\,43's SFR was calculated to be $17.72 \pm 0.03 \times 10^{-4}$ M$_\odot\textrm{yr}^{-1}$ with a conversion factor of 4.42 from FUV luminosity \cite{Subramanian_2024}.

DDO\,43 possesses an extended HI-envelope. It is an isolated galaxy with the closest object, another dwarf galaxy, being $270$\,kpc apart. It has a Holmberg-radius (where the surface brightness is 26.5 mag arcsec$^{-2}$) of 1.4 kpc in B band \cite{Simpson_2005}.

\section{Materials and methods}
\label{sec:method}

\subsection{The processing of HI spectra}

DDO\,43 was measured by the Karl G. Jansky Very Large Array (VLA) radio telescope, and its data was processed in the Astronomical Image Processing System (AIPS) during the LITTLE THINGS survey. The observational setup and data processing is described in detail in \cite{Hunter_2012}. Fully calibrated and combined data cubes are available to download on the LITTLE THINGS website \footnote{https://science.nrao.edu/science/surveys/littlethings/data/d43.html. Data available under VLA HI: Observations. Last access: 2024-05-21}. The following analysis is based on a natural-weighted data cube of DDO\,43 \cite{Hunter_2012}. 

HI emission line profiles obtained from the data cube were fit with Gaussian curves in the “Continuum and Line Analysis Single-Dish Software” (CLASS) package. CLASS is part of the “Grenoble Image and Line Data Analysis Software” (GILDAS)\footnote{http://www.iram.fr/IRAMFR/GILDAS. Last access: 2024-05-21} package, a software for processing and analysing radio-astronomical observations. The FITS data cube of DDO\,43 was converted adequately to the required input format for GILDAS CLASS.

\begin{figure*}
	\centering
	   \includegraphics[width=0.75\linewidth]{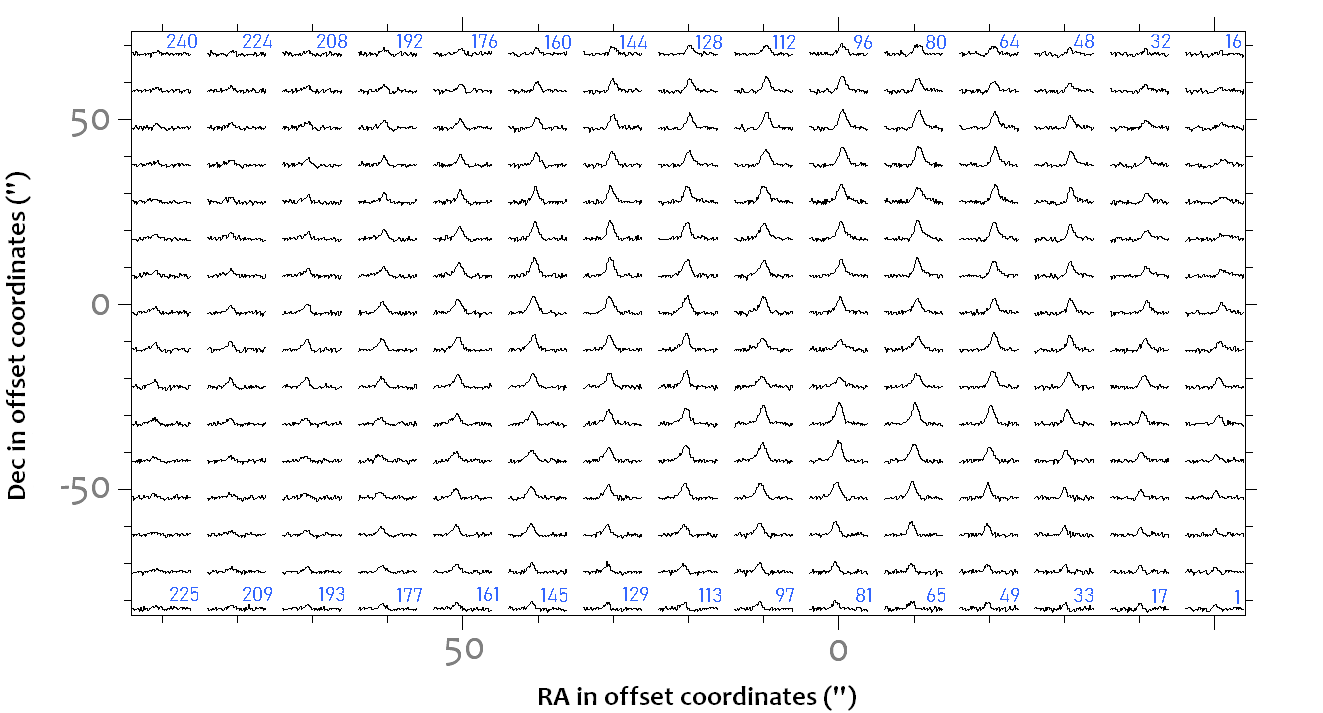}
	\hspace{5pt}
\caption{The spectra map of the analysed area, consisting of $240$ averaged spaxels. The axes are $\Delta RA$ and $\Delta Dec$ offset coordinates in arcseconds. Offset 0;0 corresponds to the galaxy's centre:  $RA= 07^\textrm{h}28^\textrm{m}17.724^\textrm{s}$, and $Dec=+40$\,\degree $46'11.36''$. The observation numbers are written in blue. Velocity and intensity scales are $274$ to $434$\,km\,s$^{-1}$ and $-0.002$ to $0.007$\,Jy\,beam$^{-1}$, respectively, see also Figure~\ref{fig:fit_spectra}.}
	\label{fig:spectramap}
\end{figure*}

\begin{figure}
	\centering
        \includegraphics[width=\linewidth]{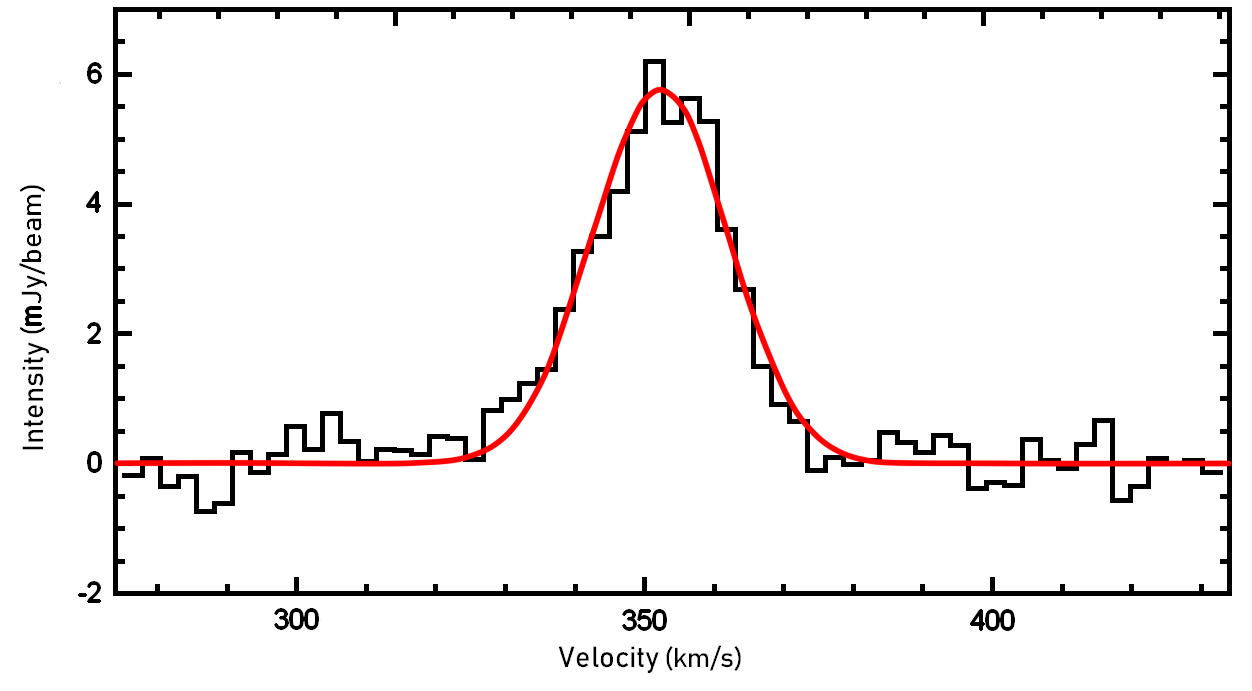}
	\\[1mm]
        \includegraphics[width=\linewidth]{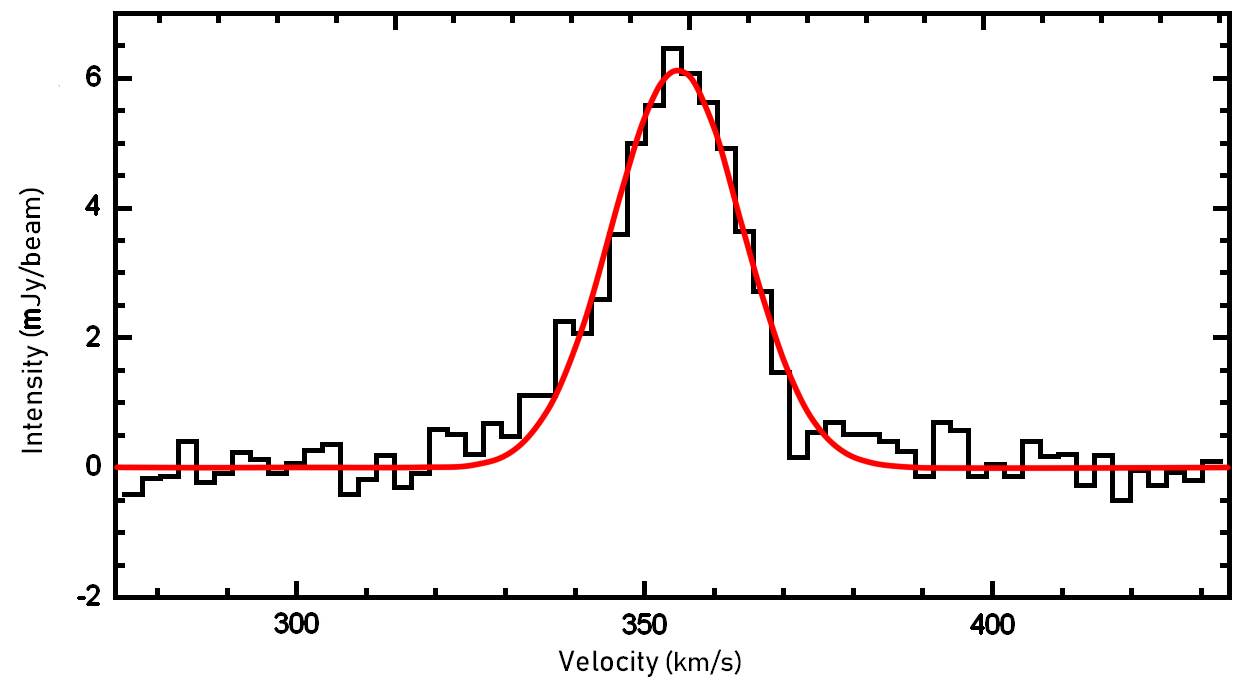}
	\caption{Example of continuum-subtracted and averaged spectra of DDO\,43. Spectra \#85 and \#86 were taken at the offset positions 0";-40" and 0";-30", respectively, (see also the spectrum map in Figure~\ref{fig:spectramap}). Red lines indicate the Gaussian fit to data.}
	\label{fig:fit_spectra}
\end{figure}

Our method for the analysis of the HI data was the following,
   \begin{enumerate}
        \item Continuum subtraction from all spectra. We used a simple first-order polynomial fit. Judging by visual inspection, this could adequately describe the continuum at any point.
        \item Averaging spectra. The whole data cube consist of $1024 \times 1024$ spectral pixels (spaxels). We determined visually from the HI line image of the galaxy that only $280 \times 280$ spaxels contain usable signal. Thus, we averaged the spectra only in this area, in radii of $5$ arcseconds, to improve the signal-to-noise ratio, resulting in $240$ spectra at the end. In Figure \ref{fig:spectramap}, we show the $240$ continuum-subtracted, averaged spectra by their offset coordinates.
        \item Automatic Gaussian fit to all spectral lines. All parameters are adjustable. CLASS describes “optimistic fits” when the base root mean square is at least $1.5$ times higher than the root mean square of the residuals on the line range(s). Example fits are shown in Figure \ref{fig:fit_spectra}: these instances are one of the strongest detections of the line, situated near the centre of the galaxy.
        \item Obtaining the residual spectra by subtracting the fits. 
   \end{enumerate}

\subsection{Far-ultraviolet data}

In order to draw a connection between the enhanced velocity dispersion and star formation, we use the Galaxy Evolution Explorer (GALEX) \cite{Martin_2005} far-ultraviolet (hereafter FUV) image of DDO\,43 \cite{Hunter_2010} for the comparison. Studying DDO\,43 in the FUV reveals where we expect the highest amount of feedback because of the short lifespan of high-mass stars and their supernovae.

The FUV flux is as observed, i.e. not corrected for dust attenuation.

\section{Results}
\label{sect:res}

The residual spectrum is the difference between the original spectral line and the fitted Gaussian curve. As the fits depend on the intensity and half-width of the line, the line wings will be mostly outside the fit. 

The residual spectra were inspected in two fixed velocity ranges. Judging by a careful visual inspection of the spectra, the furthermost extent of the line wings were $314 \textrm{\,km\,s}^{-1}$ and $394 \textrm{\,km\,s}^{-1}$ for both sides, so we identified areas that contained only visually noticeable wing features and not the full line: $(314 - 334) \textrm{\,km\,s}^{-1}$ and $(374 - 394) \textrm{\,km\,s}^{-1}$.

\begin{figure*}
	\centering
		\includegraphics[width=0.45\linewidth,height=110px]{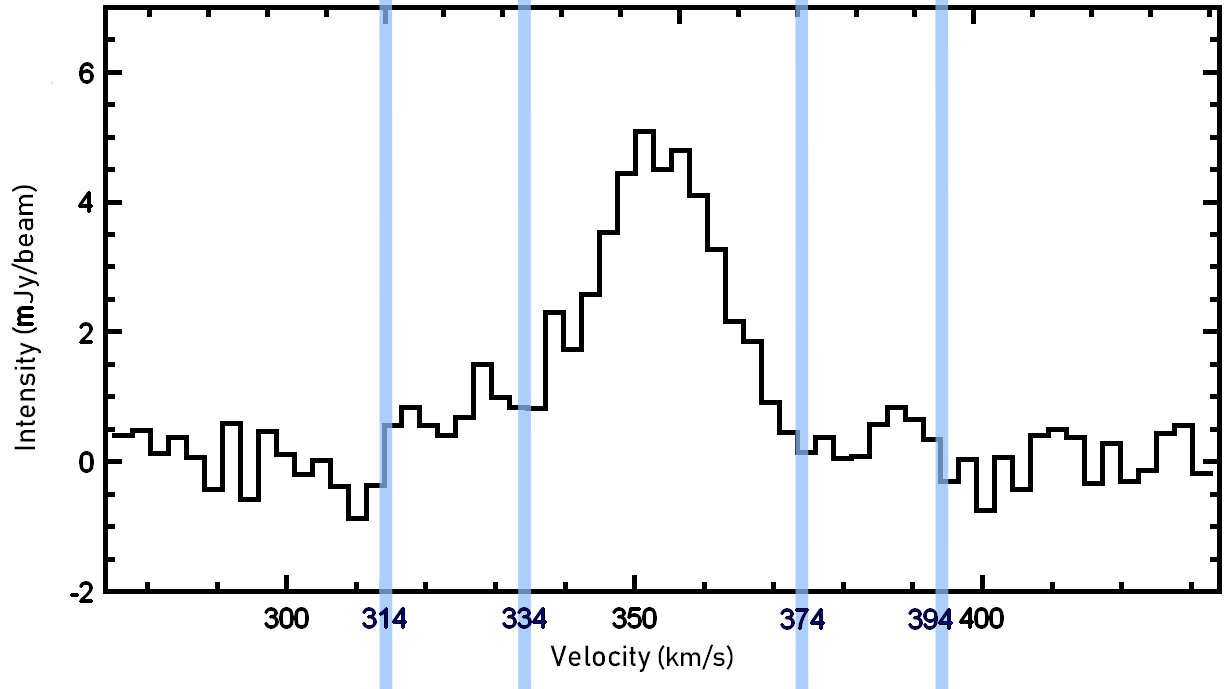}
	\hspace{5pt}
		\includegraphics[width=0.45\linewidth,height=110px]{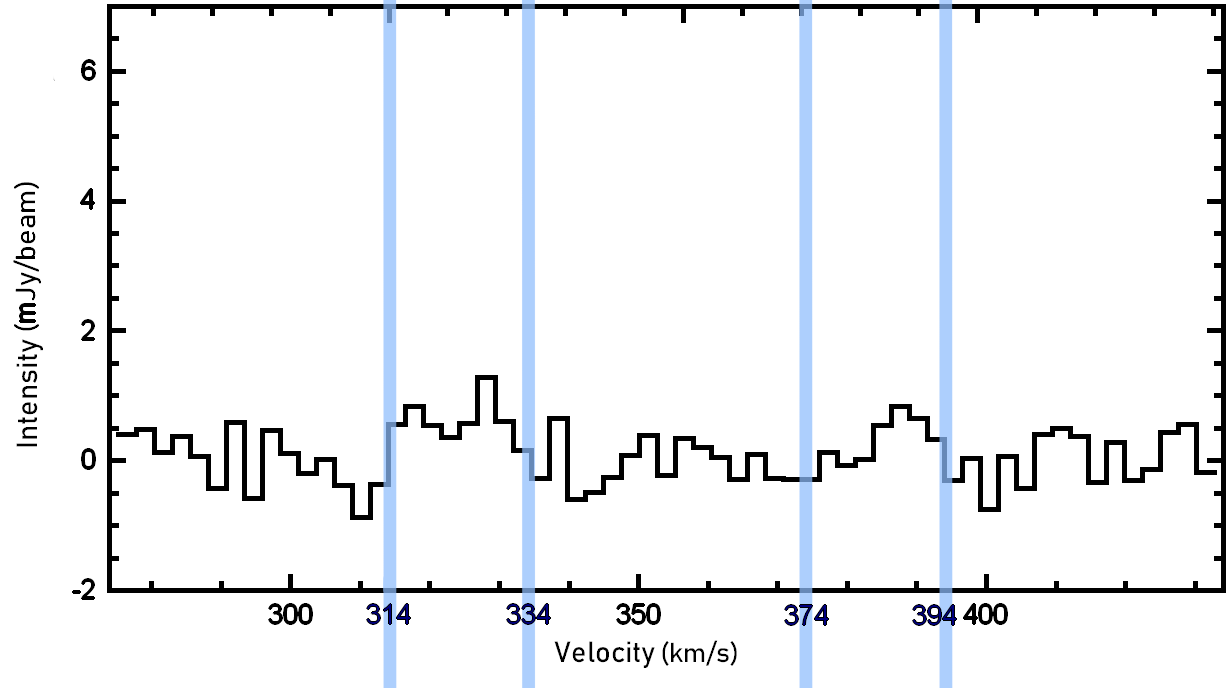}
	\caption{Left side (blue) line wing and its residual spectrum at position 105 at offset 10'', 0'' (see Figure \ref{fig:spectramap}). The blue lines represent the velocity ranges where the residual spectral line areas are inspected.}
	\label{fig:example-186}
\end{figure*}

\begin{figure*}
	\centering
		\includegraphics[width=0.45\linewidth,height=110px]{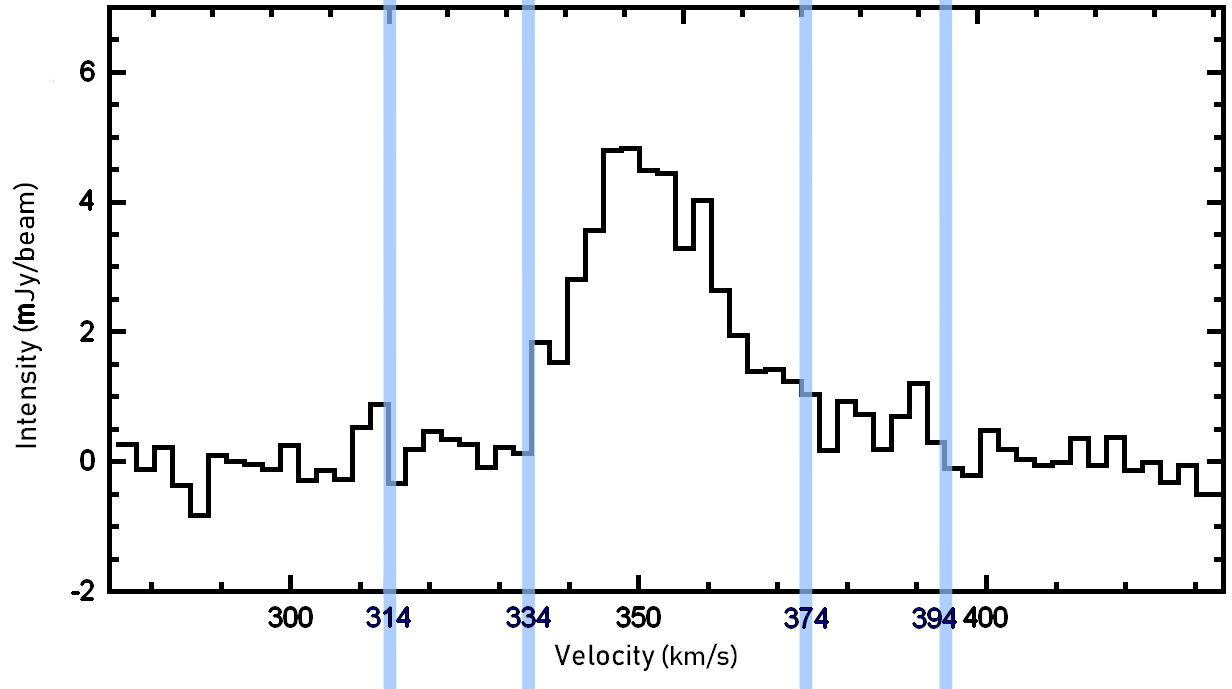}
	\hspace{5pt}
		\includegraphics[width=0.45\linewidth,height=110px]{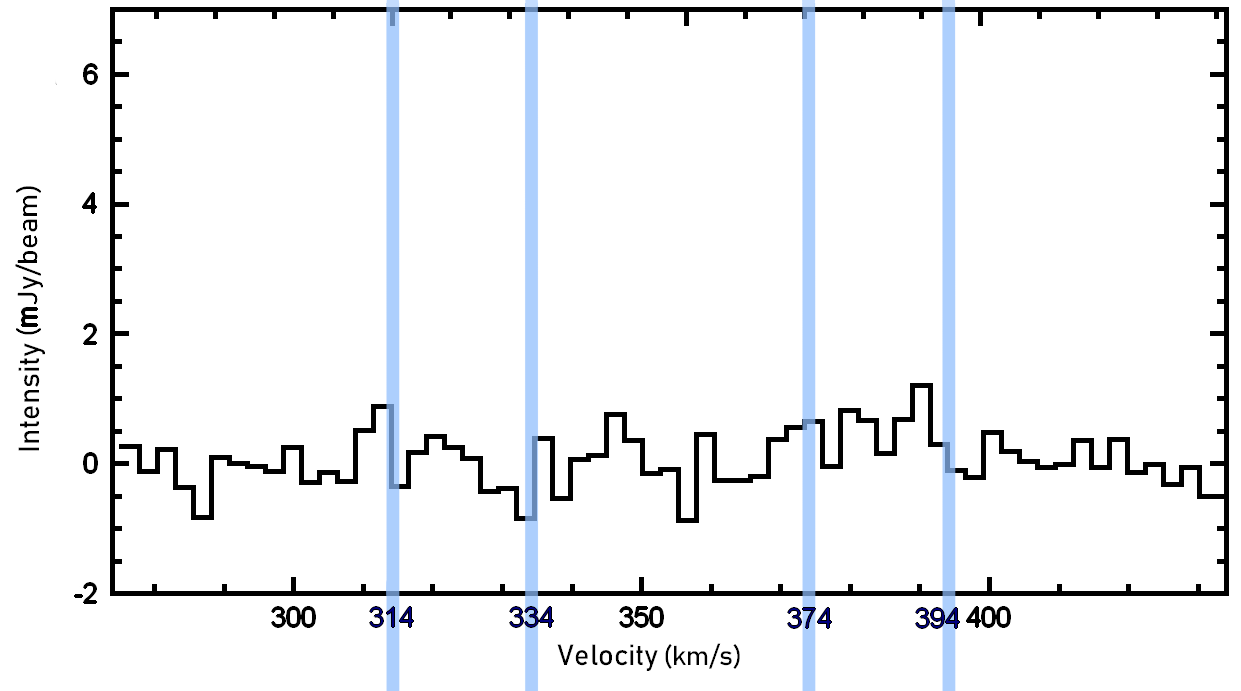}
	\caption{Right side (red) line wing and its residual spectrum at position 124 at offset 20'', 30'' (see Figure \ref{fig:spectramap}). The blue lines represent the velocity ranges where the residual spectra line areas are inspected. }
	\label{fig:example-73}
\end{figure*}

These ranges are displayed in Figures \ref{fig:example-186} and \ref{fig:example-73}, with examples of line broadening on the 'left' (blue) and on the 'right' (red) side. These examples show one of the most prominent wing features.

Table \ref{tab:area} shows the integrated residual spectral line fluxes in the noted velocity ranges.

\subsection{The distribution of residual spectral line areas}

In the case of the three spectra \#16th, \#224th, and \#227th, automatic fitting found no spectral line to fit, so these are omitted from the analysis. The average root mean square noise ($\sigma$) is $3.436 \times 10^{-4}$ Jy beam$^{-1}$. From the total 237 spectra, only whose integrated residual line fluxes were higher than 3 $\sigma$ $\times$ the channel width, $1.3 \textrm{\,km\,s}^{-1}$, have been analysed.

\begin{figure}
\centering
\includegraphics[width=\linewidth]{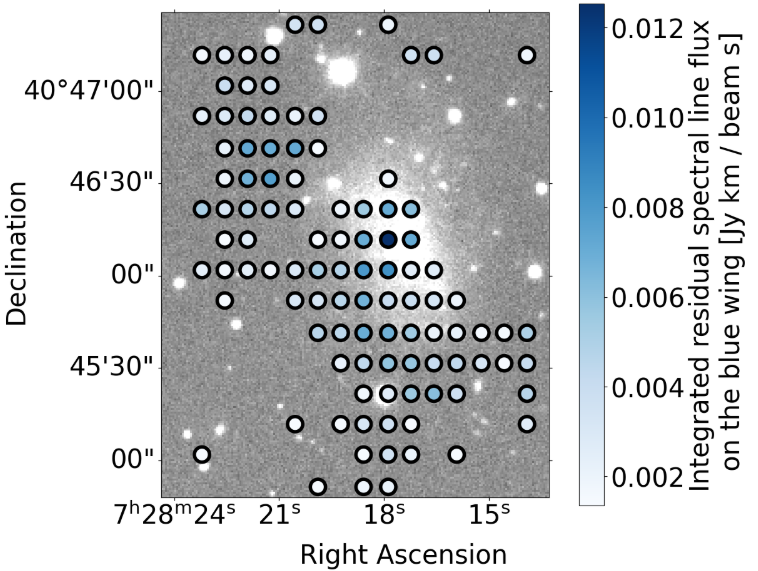}
\\[1mm]
\includegraphics[width=\linewidth]{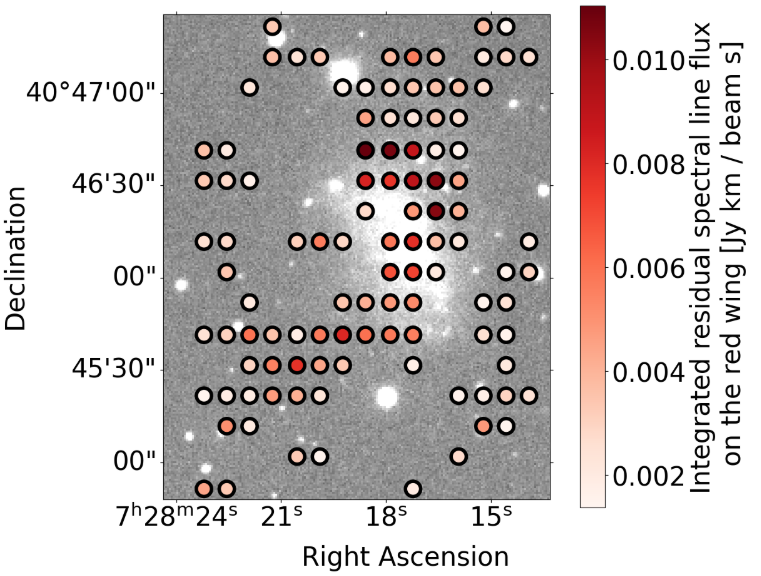}
\caption{HI line area distribution maps in velocity intervals $314 - 334 \textrm{\,km\,s}^{-1}$ (top) and $374 - 394 \textrm{\,km\,s}^{-1}$ (bottom). Each circle represents a spectrum, and the colour of the circles represents the line area: darker circles have larger line areas. The background is the optical image of the galaxy from the Sloan Digital Sky Survey \cite{Baillard_2011}.}
\label{fig:area}
\end{figure}

The line areas in the determined ranges were placed on the optical image of DDO\,43 (Sloan Digital Sky Survey, \cite{Baillard_2011}), as seen in Figure \ref{fig:area}. The alignment of the plots is the same as in the spectra map (\ref{fig:spectramap}).

As seen in Figure \ref{fig:area}, the areas with different velocity ranges are clearly separated. The velocity values change smoothly across the ranges. The areas with velocity range between $(314 - 334)\textrm{\,km\,s}^{-1}$ are in the centre and the south-western part of the galaxy, and extends to the north-east and south, which have no optical counterpart. Positions with a velocity range between $(374 - 394) \textrm{\,km\,s}^{-1}$ are on the northern part and extend from the centre to the south-east. There are no traces of these motions to the north-east.

\subsection{Comparison to far-ultraviolet data}

By analysing the residual spectral line fluxes, which provide information on the enhanced velocity dispersion in HI, our results were compared with those of the GALEX FUV image of DDO\,43 (see Section \ref{sec:method}) to investigate the relationship between the velocity dispersion and areas of massive star formation.

\cite{Subramanian_2024} examined star-forming clumps in the galaxy, and calculated the maximum star formation rate density of the clumps to be $10.2 \times 10^{-4}$ \,M$_\odot$ \,yr$^{-1}$\,kpc$^{-2}$ from GALEX FUV flux, and the minimum to be $1.64 \times 10^{-4}$ \,M$_\odot$ \,yr$^{-1}$\,kpc$^{-2}$. 

\begin{figure}
\centering
\includegraphics[width=0.75\linewidth]{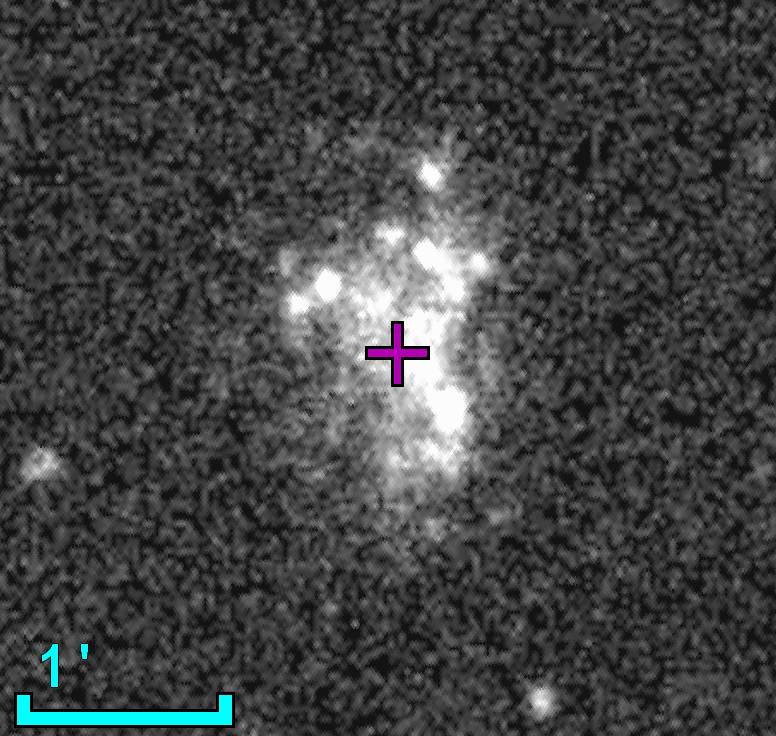}
\caption{GALEX FUV image of DDO\,43 with an effective wavelength of 151.6nm \cite{Hunter_2010}. The purple cross indicates the centre of the galaxy.}
\label{fig:uv}
\end{figure}

We looked for a correlation between the HI integrated residual spectral line fluxes, which indicate potential enhanced motion in the atomic gas, and far-ultraviolet flux, which traces massive star formation regions. A strong correlation would mean that the expected turbulence is observable in the HI gas in the regions of massive star formation, as \cite{Hunter_2021} and \cite{Elmegreen_2022} predicted. 

\begin{figure}
\centering
\includegraphics[width=\linewidth]{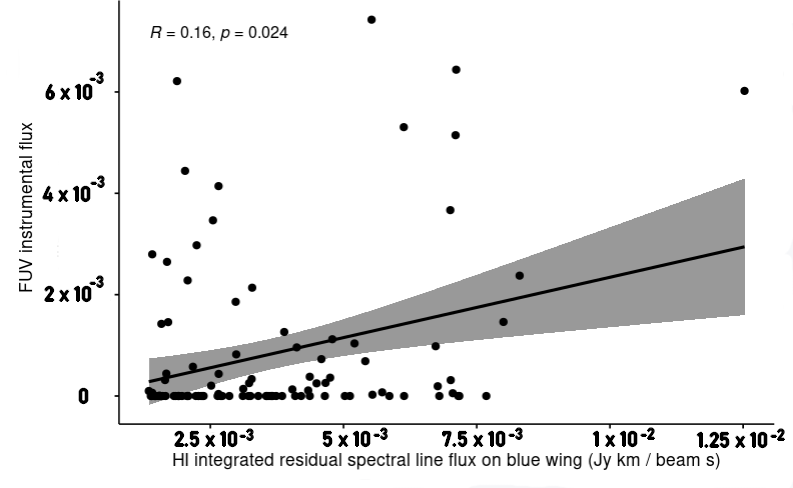}
\\[3mm]
\includegraphics[width=\linewidth]{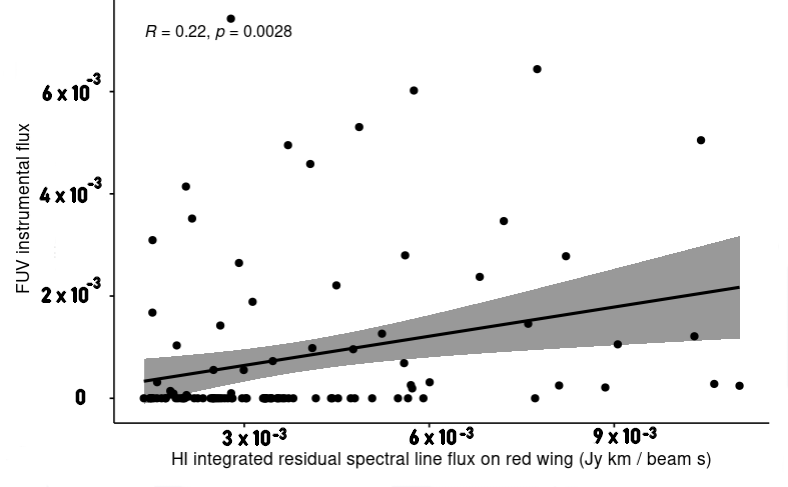}
\caption{HI line areas in the velocity intervals $314 - 334 \textrm{\,km\,s}^{-1}$ (top) and $374 - 394 \textrm{\,km\,s}^{-1}$ (bottom) against FUV flux. Between the two variables, a Kendall correlation coefficient of $0.16$ and $0.22$ can be seen, respectively.}
\label{fig:corr}
\end{figure}

To investigate this, the FUV flux was calculated from the GALEX image (Figure \ref{fig:uv}) in the positions of the spectra in Figure \ref{fig:spectramap}. We used the astropy package \cite{astropy:2013, astropy:2018, astropy:2022} to find the coordinates in the FUV image corresponding to each spectra, and then calculated the average intensity in a radius of 5 pixels around the found positions. 5 pixels correspond to the area in which the spectra were averaged and matched according to the difference between the HI image resolution and FUV resolution. We calculated  Kendall's $\tau$, which is a measure of ordinal relationship between two variables, for the HI residual spectral line flux and FUV flux pairs. The correlation plots are seen in Figure \ref{fig:corr}.

\section{Discussion}

We compared 237 blue, $(314 - 334)$\,km\,s$^{-1}$, and red, $(374 - 394)$\,km\,s$^{-1}$, integrated residual spectral line fluxes to the FUV flux. We found the following relationship between the values: 

\begin{itemize}
    \item There appears to be a very weak correlation between HI blue integrated residual spectral line fluxes and FUV fluxes, with a Kendall correlation coefficient of $0.16$ and p-value (significance) of 0.024.
    \item There is also a weak correlation between red integrated residual spectral line fluxes and FUV fluxes, with a Kendall correlation coefficient of $0.22$ and p-value of 0.0028.
\end{itemize}

With the result of a weak correlation with both line wings, our conclusion is consistent with previous studies (\cite{Hunter_2021}, \cite{Elmegreen_2022}). The expected enhanced turbulence does not seem to be observable with our method. 

Dust attenuation is generally expected to be moderate for the almost face-on galaxy, except the most central regions. There, the extinction may lower the observed FUV flux, flattening the UV vs. HI curve, thus lowering the correlation coefficient as well.

However, as our study does not take the position of the spectral lines into consideration - i.e. the rotation - with the fixed velocity ranges in which the residual spectra are measured, our method can be improved if instead of residual spectra, we fit the line wings as separate Gaussian curves, and examine them individually. 

\begin{table*}
	\begin{tabular}{rrrrr}
		\hline
		\textbf{Spectrum} & \textbf{$\Delta RA$}  & \textbf{$\Delta Dec$} & \textbf{$W_B$} & \textbf{$W_R$} \\
&"&"&Jykms$^{-1}$beam$^{-1}$&Jykms$^{-1}$beam$^{-1}$\\
		\hline
        105 & $10.0$ & $0.0$&$1.25266\times10^{-2}$ & $5.75052\times10^{-3}$  \\
		\hline
		124 &  $20.0$ & $30.0$ & $-2.30867\times10^{-3}$ & $1.10328\times10^{-2}$\\
		\hline
        85 & 0.0& -40.0& $5.72505\times10^{-3}$ & $1.81162\times10^{-3}$\\
        \hline
		86 &0.0 & -30.0& $5.40835\times10^{-3}$& $5.59232\times10^{-3}$ \\
        \hline
	\end{tabular}
	\caption{Offset coordinates and integrated residual spectral line fluxes of the spectra shown on Figures \ref{fig:example-186} and \ref{fig:example-73}, in the ranges marked with blue, and also of the spectra shown in Figure \ref{fig:fit_spectra} in the respective ranges. The average error of area calculation in CLASS was $1.79 \times 10^{-3}$.}
	\label{tab:area}
\end{table*}

\section{Conclusions}

We investigated the relationship between star formation parameters and interstellar gas in the dwarf galaxy DDO\,43. We looked for an apparent broadening of the HI spectral lines by inspecting their residual spectra. 

We obtained integrated HI residual spectral line fluxes indicating enhanced velocity dispersion at $237$ positions of DDO\,43, which are plotted on distribution maps throughout the galaxy. 

Comparing the integrated residual spectral line fluxes to FUV flux, we find very weak correlations with a Kendall-coefficient of 0.16 on the blue side and 0.22 on the red side.

\begin{acknowledgements}

We thank the anonymous reviewers for the careful reading of the manuscript and their valuable comments and suggestions.

E. Pichler would like to thank Dr. Deidre Hunter and Dr. Bruce Elmegreen for their consultations regarding DDO\,43 data and line wings research, and
S\'ebastien Bardeau for providing help with the use of the CLASS software.

The IBWS conference participation of B. Koncz was subsidised by the Dean's Council of ELTE Eötvös Loránd University Faculty of Science, Budapest. 

K\'EG received funding from the Hungarian National Research, Development and Innovation Office (NKIFH) through the grant OTKA K134213, and the NKFIH excellence grant TKP2021-NKTA-64. K\'EG also acknowledges the support received from the HUN-REN Hungarian Research Network. 

The National Radio Astronomy Observatory is a facility of the National Science Foundation operated under cooperative agreement by Associated Universities, Inc.

This research has made use of the NASA/IPAC Extragalactic Database (NED), which is funded by the National Aeronautics and Space Administration and operated by the California Institute of Technology.

This research has made use of “Aladin sky atlas” developed at CDS, Strasbourg Observatory, France. 

\end{acknowledgements}

\bibliographystyle{actapoly}
\bibliography{biblio}

\begin{thebibliography}{10}
\providecommand{\url}[1]{\texttt{#1}}
\providecommand{\urlprefix}{URL }
\providecommand{\eprint}[2][]{\url{#2}}
\makeatletter
\def\bibdoi{\begingroup\def\do##1{\catcode
  `##112\relax}\do$\do\\\do\_\do\%\do\^\expandafter\endgroup\@bibdoi}
\def\@bibdoi#1{\eprint{https://doi.org/#1}}
\makeatother

\bibitem{Law_2022}
D.~R. Law, F.~Belfiore, M.~A. Bershady, et~al.
\newblock Sdss-iv manga: Understanding ionized gas turbulence using integral
  field spectroscopy of 4500 star-forming disk galaxies.
\newblock \textit{The Astrophysical Journal} \textbf{928}(1):58, 2022.
\bibdoi{10.3847/1538-4357/ac5620}
\bibitem{Bacchini_2020}
{Bacchini, Cecilia}, {Fraternali, Filippo}, {Iorio, Giuliano}, et~al.
\newblock Evidence for supernova feedback sustaining gas turbulence in nearby
  star-forming galaxies.
\newblock \textit{A\&A} \textbf{641}:A70, 2020.
\bibdoi{10.1051/0004-6361/202038223}
\bibitem{Elmegreen_2004}
B.~G. Elmegreen, J.~Scalo.
\newblock Interstellar turbulence i: Observations and processes.
\newblock \textit{Annual Review of Astronomy and Astrophysics}
  \textbf{42}(Volume 42, 2004):211--273, 2004.
\bibdoi{https://doi.org/10.1146/annurev.astro.41.011802.094859}
\bibitem{Ianjamasimanana_2015}
R.~Ianjamasimanana, W.~J.~G. de~Blok, F.~Walter, et~al.
\newblock The radial variation of h i velocity dispersions in dwarfs and
  spirals.
\newblock \textit{The Astronomical Journal} \textbf{150}(2):47, 2015.
\bibdoi{10.1088/0004-6256/150/2/47}
\bibitem{Elmegreen_2022}
B.~G. Elmegreen, Z.~Martinez, D.~A. Hunter.
\newblock A search for correlations between turbulence and star formation in
  things galaxies.
\newblock \textit{The Astrophysical Journal} \textbf{928}(2):143, 2022.
\bibdoi{10.3847/1538-4357/ac559c}
\bibitem{Hunter_2021}
D.~A. Hunter, B.~G. Elmegreen, H.~Archer, et~al.
\newblock A search for correlations between turbulence and star formation in
  little things dwarf irregular galaxies.
\newblock \textit{The Astronomical Journal} \textbf{161}(4):175, 2021.
\bibdoi{10.3847/1538-3881/abe1c0}
\bibitem{Karachentsev_2004}
I.~D. Karachentsev, V.~E. Karachentseva, W.~K. Huchtmeier, D.~I. Makarov.
\newblock A catalog of neighboring galaxies.
\newblock \textit{The Astronomical Journal} \textbf{127}(4):2031, 2004.
\bibdoi{10.1086/382905}
\bibitem{Hunter_2012}
D.~A. Hunter, D.~Ficut-Vicas, T.~Ashley, et~al.
\newblock Little things.
\newblock \textit{The Astronomical Journal} \textbf{144}(5):134, 2012.
\bibdoi{10.1088/0004-6256/144/5/134}
\bibitem{Simpson_2005}
C.~E. Simpson, D.~A. Hunter, T.~E. Nordgren.
\newblock Ddo 43: A prototypical dwarf irregular galaxy?
\newblock \textit{The Astronomical Journal} \textbf{130}(3):1049, 2005.
\bibdoi{10.1086/432537}
\bibitem{chambers2019panstarrs1surveys}
K.~C. Chambers, E.~A. Magnier, N.~Metcalfe, et~al.
\newblock The pan-starrs1 surveys, 2019.
\eprint{1612.05560}
\bibitem{Flewelling_2020}
H.~A. Flewelling, E.~A. Magnier, K.~C. Chambers, et~al.
\newblock The pan-starrs1 database and data products.
\newblock \textit{The Astrophysical Journal Supplement Series}
  \textbf{251}(1):7, 2020.
\bibdoi{10.3847/1538-4365/abb82d}
\bibitem{Hunter_2004}
D.~A. Hunter, B.~G. Elmegreen.
\newblock Star formation properties of a large sample of irregular galaxies.
\newblock \textit{The Astronomical Journal} \textbf{128}(5):2170, 2004.
\bibdoi{10.1086/424615}
\bibitem{Subramanian_2024}
{Subramanian, Smitha}, {Mondal, Chayan}, {Kalari, Venu}.
\newblock Effect of low-mass galaxy interactions on their star formation.
\newblock \textit{A\&A} \textbf{681}:A8, 2024.
\bibdoi{10.1051/0004-6361/202346536}
\bibitem{Martin_2005}
D.~C. Martin, J.~Fanson, D.~Schiminovich, et~al.
\newblock The galaxy evolution explorer: A space ultraviolet survey mission.
\newblock \textit{The Astrophysical Journal} \textbf{619}(1):L1, 2005.
\bibdoi{10.1086/426387}
\bibitem{Hunter_2010}
D.~A. Hunter, B.~G. Elmegreen, B.~C. Ludka.
\newblock Galex ultraviolet imaging of dwarf galaxies and star formation
  rates*.
\newblock \textit{The Astronomical Journal} \textbf{139}(2):447, 2010.
\bibdoi{10.1088/0004-6256/139/2/447}
\bibitem{Baillard_2011}
A.~Baillard, E.~Bertin, V.~de~Lapparent, et~al.
\newblock The efigi catalogue of 4458 nearby galaxies with detailed morphology.
\newblock \textit{Astronomy \& Astrophysics} \textbf{532}:A74, 2011.
\bibdoi{10.1051/0004-6361/201016423}
\bibitem{astropy:2013}
{Astropy Collaboration}, T.~P. {Robitaille}, E.~J. {Tollerud}, et~al.
\newblock {Astropy: A community Python package for astronomy}.
\newblock \textit{\aap} \textbf{558}:A33, 2013.
\eprint{1307.6212} \bibdoi{10.1051/0004-6361/201322068}
\bibitem{astropy:2018}
{Astropy Collaboration}, A.~M. {Price-Whelan}, B.~M. {Sip{\H{o}}cz}, et~al.
\newblock {The Astropy Project: Building an Open-science Project and Status of
  the v2.0 Core Package}.
\newblock \textit{\aj} \textbf{156}(3):123, 2018.
\eprint{1801.02634} \bibdoi{10.3847/1538-3881/aabc4f}
\bibitem{astropy:2022}
{Astropy Collaboration}, A.~M. {Price-Whelan}, P.~L. {Lim}, et~al.
\newblock {The Astropy Project: Sustaining and Growing a Community-oriented
  Open-source Project and the Latest Major Release (v5.0) of the Core Package}.
\newblock \textit{\apj} \textbf{935}(2):167, 2022.
\eprint{2206.14220} \bibdoi{10.3847/1538-4357/ac7c74}
\end{thebibliography}

\clearpage
\appendix
\onecolumn
%\section{Appendix}
%\label{sec:appx}

%\subsection{External data sets}
%Use this first Appendix for description and link to external datasets.

%\subsection{Large data sets}
%Any supplementary large sets of data (e.g. graphs, tables) that are not appropriate to insert into the text.

\end{document}